\documentclass[12pt]{article}

\usepackage{amsmath,amsthm,amsfonts,amssymb}
\usepackage{graphics,psfrag}
\usepackage{cite}
\usepackage{array,epsfig,multirow,stmaryrd,graphicx}
\usepackage{hyperref}
\usepackage[utf8]{inputenc}
\usepackage[mathscr]{eucal}
\usepackage{graphicx}
\usepackage{color}
\usepackage{hypbmsec}
\usepackage{xspace}
\usepackage{rotating}
\usepackage[vcentermath]{youngtab}
\usepackage[isu,bf]{caption}
\def\hybrid{\topmargin -20pt    \oddsidemargin 0pt
        \headheight 0pt \headsep 0pt 
        \textwidth 6.25in      
        \textheight 9 in      
        \marginparwidth .875in
        \parskip 5pt plus 1pt
          \jot = 1.5ex
  }
\hybrid
\numberwithin{equation}{section}
\numberwithin{table}{section}

\newcommand{\cC}{\mathcal{C}}

\newcommand{\rep}[1]{\mathbf{#1}}

\newcommand{\cref}{{\bf [check ref]}}

\DeclareMathOperator{\sign}{sign}

\newcommand{\Q}{\ensuremath{{\mathbb{Q}}}}

\newcommand{\C}{\ensuremath{{\mathbb{C}}}}

\newcommand{\Z}{\mathbb{Z}}
\newcommand{\CP}{{\ensuremath{\mathop{\null {\mathbb{P}}}\nolimits}}}

\DeclareMathOperator{\Id}{id}
\DeclareMathOperator{\rank}{rank}

\newcommand{\Rep}[1]{\mathbf{#1}}

\newcommand{\Osheaf}{\ensuremath{\mathscr{O}}}

\newcommand{\tmod}{~\mathrm{mod}~}

\newenvironment{descriptionlist}{%
\begin{list}%
{}%
{}}%
{\end{list}}

{\everymath{\displaystyle\everymath{}}\array}%
{\endarray}

\newcommand{\Dhat}[1]{\ensuremath{\hat{D}_{#1}}}
\newcommand{\Hhat}[1]{\ensuremath{\hat{H}_{#1}}}
\newcommand{\Fhat}[1]{\ensuremath{\hat{F}_{#1}}}

\newcommand{\cDbar}[1]{\ensuremath{\bar{D}_{#1}}}
\newcommand{\cFbar}[1]{\ensuremath{\bar{F}_{#1}}}

\begin{document}

\baselineskip=15pt

\begin{titlepage}
\begin{flushright}
\parbox[t]{1.08in}{MPP-2013-27}
\end{flushright}

\begin{center}
\vspace*{ 1.2cm}
{\large \bf New Global F-theory GUTs with U(1) symmetries}
\vskip 1.2cm

\renewcommand{\thefootnote}{}
\begin{center}
 {Volker Braun$\,^1$, Thomas W.~Grimm$\, ^2$ and Jan Keitel$\,^2$ 
   \footnote{\texttt{vbraun@stp.dias.ie},\ \texttt{grimm}, \texttt{jkeitel@mpp.mpg.de}}}
\end{center}
\vskip .2cm
\renewcommand{\thefootnote}{\arabic{footnote}} 

{$\,^1$ Dublin Institute for Advanced Studies,\\
10 Burlington Road, Dublin 4, Ireland}

{$\,^2$ Max-Planck-Institut f\"ur Physik, \\
F\"ohringer Ring 6, 80805 Munich, Germany}

 \vspace*{1cm}

\end{center}

\vskip 0.2cm
 
\begin{center} {\bf ABSTRACT } \end{center} 

\noindent We construct global F-theory GUTs with $SU(5) \times U(1)$
gauge group defined by specifying a fully resolved Calabi-Yau fourfold
and consistent four-form G-flux. Its specific $U(1)$ charged matter
spectrum allows the desired Yukawa couplings, but forbids dangerous
proton decay operators.  The model we find: (1) does not follow from
an underlying higgsed $E_8$ gauge group (2) leaves the class of
theories that can be analyzed with current split-spectral cover
techniques. This avoids recently proposed no-go theorems for models
with hypercharge flux, as required to break the GUT group. The
appearance of additional fields is related geometrically to
considering a more general class of sections and 4--1 splits. We show
explicitly that the four-dimensional chiral matter index can still be
computed using three-dimensional one-loop Chern-Simons terms.
 
\end{titlepage}

\section{Introduction and Summary}

Over the past years, considerable effort has been spent on engineering
phenomenologically viable $SU(5)$-GUTs in F-theory. This was initiated
in local models with decoupled gravity in~\cite{Donagi:2008ca,
  Beasley:2008dc, Beasley:2008kw}. Since then, vast progress has been
made both in providing a more detailed local construction as well as
in finding global completions~\cite{Weigand:2010wm,Maharana:2012tu}.
Despite various successes such as the construction of fully resolved
fourfolds with $SU(5)$ and $SO(10)$ gauge group
singularities~\cite{Blumenhagen:2009yv,Grimm:2009yu,Chen:2010ts,
  Knapp:2011wk,Marsano:2011hv}, fully consistent phenomenological
models have yet to be found. This can, at least in a large class of
GUT scenarios, be traced back to the fact that $U(1)$
symmetries~\cite{Morrison:1996pp, Aspinwall:1998xj,
  Aspinwall:2000kf,Grimm:2010ez, Morrison:2012ei,Mayrhofer:2012zy} are
difficult to control in F-theory. This is naturally so because they,
unlike non-Abelian gauge factors, do not localize and depend on the
global properties of the Calabi-Yau manifold. Moreover, Abelian gauge
factors can be crucial in order to prohibit proton decay in GUT
models. That is, appropriate $U(1)$-charges can be used to allow only
the desired couplings and forbid proton decay inducing
couplings. However, the actual $U(1)$ gauge group (and, therefore,
charges) depend on the global Calabi-Yau geometry and are difficult to
engineer.

The authors of~\cite{Marsano:2009gv, Marsano:2009wr, Dudas:2010zb,
  Dolan:2011iu} developed local methods in order to classify and study
F-theory seven-brane gauge theories with gauge group $SU(5) \times
U(1)^k$. In this framework, the 7-brane theory is determined by an
auxiliary geometric object, the spectral cover $\cC$, which encodes
information about a Higgs fields used to break an underlying $E_8$
symmetry as
\begin{equation}
   E_8 \longrightarrow 
   SU(5)_{GUT} \times SU(5)_{\perp} \longrightarrow
   SU(5)_{GUT} \times U(1)^4 .
\end{equation}
Matter states in the 7-brane worldvolume theory originate from the
decomposition of the $\rep{248}$ representation of $E_8$. In the
general case, monodromies of the spectral cover then lead to further
identifications among some of the Abelian gauge group factors. More
precisely, in the presence of a $U(1)^k$ gauge group factor, the
spectral cover splits into $k+1$ irreducible components. The structure
of $E_8$ determines that $\cC$ is defined by a degree five polynomial
and therefore $k \leq 4$. Accordingly, there are exactly two
possibilities in the case of a single $U(1)$, namely the factorization
into a linear and a quartic factor and the factorization into a
quadratic and a cubic piece. These two cases are dubbed 4--1 split and
3--2 split, respectively, and their associated field contents are
displayed in \autoref{t:splits}
\begin{table}[h!]
  \centering
  \begin{tabular}{|c|c|c|}
    \hline
    Model & \multicolumn{2}{c|}{Matter spectrum}\\
    \hline
    4--1 split & $\rep{5_{-3}}$, $\rep{5_2}$, $\rep{10_{-1}}$, $\rep{10_4}$ &  $\rep{1_5}$ \\
    3--2 split & $\rep{5_{-4}}$, $\rep{5_1}$, $\rep{5_6}$, $\rep{10_{-3}}$, $\rep{10_2}$ & $\rep{1_5}$\\
    \hline
  \end{tabular}
  \caption{Different $SU(5) \times U(1)$ spectra originating from
    $E_8$ branching rules.}
  \label{t:splits}
\end{table}
By definition, the 7-brane spectral cover captures only physics that
is localized on the discriminant and contains no information about the
existence of a $U(1)$ gauge symmetry in the low-energy effective
action or the spectrum of $SU(5)$-singlets, as has been noted various
times in the literature~\cite{Donagi:2009ra, Grimm:2010ez,
  Dolan:2011iu, Mayrhofer:2012zy}.

Globally, the appearance of extra $U(1)$ factors is equivalent to
additional sections generating a non-trivial Mordell-Weil group of the
Calabi-Yau manifold. Global models with sections coming from a
factorized Tate model where studied recently
in~\cite{Mayrhofer:2012zy}. In this paper, we construct an F-theory
GUT with $SU(5) \times U(1)$ low-energy gauge symmetry by specifying a
concrete elliptically fibered Calabi-Yau manifold that is not
contained in the above class of models. A priori there is no reason
for the spectral cover and the low-energy gauge group to match. We find the
first concrete model in the literature where such a match does not occur. The
low-energy spectrum does not fit into $\Rep{248}$ of $E_8$ due to the
$\rep{1}_{10}$ singlet states, as was first discovered
in~\cite{Mayrhofer:2012zy}. In our example, there are in addition four
distinctly charged $\rep{5}$ representations of $SU(5)$ which are also
at odds with a spontaneously broken $E_8$. Their appearance can be
explained by considering a more general class of rational
sections. Unlike their holomorphic counterparts, rational sections are
allowed to wrap entire fiber components instead of intersecting them
transversely~\cite{Morrison:2012ei, Mayrhofer:2012zy}, thereby
allowing a larger variety of $U(1)$ charges. Using the intersection
properties of the fiber geometry, we give a simple generalized
geometric interpretation of the classification into different splits
and propose that there exists another split that has not been studied
so far.

According to this generalized definition of splits, our model is an
example of a 4--1 split. However, due to the existence of four
distinct $\rep{5}$ representations instead of just two, two
phenomenologically desirable features can be realized: First of all, a
Peccei-Quinn type symmetry can be used to forbid a $\mu$-term and
dimension five proton decay operators. Second of all, our set of
matter curves admits in principle non-trivial anomaly free solutions
to the constraints imposed on hypercharge flux that were examined
in~\cite{Palti:2012dd}.

These results are presented as follows: In \autoref{s:field_theory} we present the field content of our GUT in detail, describe how to match the different MSSM fields with the different representations of $SU(5) \times U(1)$ and explain the conditions imposed by anomaly cancellation. We then proceed with a detailed construction of the Calabi-Yau manifold in terms of a hypersurface in a toric variety in \autoref{s:cy}. Next of all, the matter spectrum is analyzed and interpreted geometrically in \autoref{sec:charge}. Finally, we construct a set of consistent fourform fluxes in \autoref{s:chiralities}. By using three-dimensional one-loop Chern-Simons
terms, we compute the chiral indices of the four-dimensional matter representations and show that the spectrum is anomaly free.

\section{More general GUT Models: Field
  Theory} \label{s:field_theory}

Let us start with a brief field theoretic description of the model
that we will construct in the remainder of this paper. The
representation content of our four-dimensional theory is
\begin{align}
  \rep{10_{-1}},~ \rep{5_{-8}},~  \rep{5_{-3}},~  \rep{5_{2}},~
  \rep{5_{7}},~ \rep{1_5},~ \rep{1_{10}},
  \label{e:field_content}
\end{align}
where the conjugate representations are understood to be included as
well. Recalling that the necessary Yukawa couplings in order to
reproduce the MSSM are
\begin{align}
  \mathbf{10_M} \mathbf{10_M} \mathbf{5_H} \quad \text{and} \quad \mathbf{10_M} \mathbf{\overline{5}_M} \mathbf{\overline{5}_H},
\end{align}
we match the MSSM representations as
\begin{equation}
  \rep{\overline{5}_M} \longleftrightarrow \rep{\overline{5}_{-7}} \qquad
  \rep{5_H} \longleftrightarrow \rep{5_2} \qquad 
  \rep{\overline{5}_H} \longleftrightarrow \rep{\overline{5}_8}  \qquad
  \rep{10_M} \longleftrightarrow \rep{10_{-1}} 
  .
\end{equation}
Note that one could also swap the $U(1)$-charges assigned to
$\rep{\overline{5}_M}$ and $\rep{\overline{5}_H}$. With the above
choice, both the dimension four proton decay operator
\begin{equation}
  \rep{10_M} \rep{5_M} \rep{5_M}
\end{equation}
as well as a tree-level $\mu$-term in the superpotential
\begin{equation}
  \mu \, \rep{5_H} \rep{\overline{5}_H}
\end{equation}
are forbidden by the $U(1)$-symmetry. The absence of the latter terms
is equivalent to prohibiting operators such as
\begin{equation}
  \rep{10_M} \rep{10_M} \rep{10_M} \rep{\overline{5}_M}
  ,
\end{equation}
which would induce dimension 5 operators after integrating out heavy
modes. Let us remark that $U(1)$-symmetries with charge assignments
such that
\begin{equation}
  Q(\rep{5_H}) \neq - Q(\rep{\overline{5}_H})
\end{equation}
are called Peccei-Quinn (PQ) symmetries. The remaining representation
$\rep{5_{-3}}$ will carry exotic matter, while the $SU(5)$-singlets
are candidates for right-handed neutrinos.

Having introduced the spectrum, we now turn to the conditions any
anomaly-free spectrum must satisfy. Before turning on hypercharge flux
to break up the above $SU(5)$-multiplets, these conditions are 
\begin{equation}  \label{e:su5_anomalies}
  \begin{split}
    0=&\,   \chi(\rep{10_{-1}}) + \sum_q \chi(\rep{5_q}),  \\
    b^\alpha_{(5)} \Theta_{\alpha 5} 
    =&\, - \frac{20}{3} \chi(\rep{10_{-1}}) + \frac{10}{3} \sum_q \chi(\rep{5_q}) q^3 + \frac{2}{3} \sum_p \chi(\rep{1_p}) p^3   ,\\
    - a^\alpha \Theta_{\alpha 5} 
    =&\,- \frac{10}{3} \chi(\rep{10_{-1}}) + \frac{5}{3} \sum_q \chi(\rep{5_q}) q + \frac{1}{3} \sum_p \chi(\rep{1_p}) p  ,\\
    b^\alpha \Theta_{\alpha 5}
    =&\, 6 \chi(\rep{10_{-1}}) + 2 \sum_q \chi(\rep{5_q}) q .
  \end{split}
\end{equation}
They follow from demanding that the non-Abelian anomaly, the purely
Abelian anomaly and two mixed anomalies are cancelled,
respectively. The index $q \in \{-8,-3,2,7\}$ ranges over the
different charges of the $\rep{5}$-representations and $p$ runs over
the $U(1)$-charges of the $SU(5)$-singlets. $a^\alpha$, $b^\alpha$ and
$b^\alpha_{(5)}$ are coefficients of the Green-Schwarz terms and
$\Theta_{\alpha 5}$ is a Chern-Simons coefficient, all of which will
be related to compactification data in \autoref{ss:fourfold}. The relation between geometric data and Green-Schwarz coefficients has recently been studied in the context of anomaly cancellation in F-Theory in~\cite{Cvetic:2012xn}.

After turning on hypercharge flux an additional set of anomaly
cancellation conditions must be imposed. These can be understood as
constraining all possible hypercharge fluxes. For a single $U(1)$ in
the low-energy effective action there are four constraints,
namely~\cite{Marsano:2010sq, Palti:2012dd}
\begin{equation}
  \begin{gathered}
    N^{-1}_{10} = \sum_q N_5^q = 0, \\
    - N^{-1}_{10} + \sum_q N_5^q q = 0, \\
    3 N^{-1}_{10} + \sum_q N_5^q q^2 = 0, 
    \label{e:hyperflux_constraints}
  \end{gathered}
\end{equation}
where $N^{a}_b$ denotes the number of flux quanta along
$\rep{b_a}$. Since we have five parameters and four linear
constraints, the above equations admit a one parameter set of
solutions:
\begin{equation}
  \begin{gathered}
    N^{-1}_{10} = 0, \\
    N_5^{-8} = \lambda, \qquad N_5^{-3} = - 3 \lambda, \\
    N_5^{2} = 3 \lambda, \qquad N_5^{7} = - \lambda.
  \end{gathered}
\end{equation}

We close this section by emphasizing that the above matter spectrum
with its additional $\rep{5}$ representations might allow to
circumvent the ``no-go''-theorem forbidding hypercharge flux in models
with $4$--$1$ split~\cite{Palti:2012dd}. To be precise, the anomaly
constraints \eqref{e:hyperflux_constraints} and their generalized
version for cases with several distinctly charged $\rep{10}$
representations always eliminate precisely four flux parameters in
models with a single $U(1)$ factor, as we see above. The
``no-go''-theorem is based on a counting argument~\cite{Palti:2012dd}:
Since the $4$--$1$ splits in models with an underlying $E_8$ symmetry
never have more than four distinct matter curves charged under
$SU(5)$, the only solution to the anomaly equations in these models is
to have no flux. However, we will show in the next section that there
indeed exist models of a very similar type that can have more
representations and therefore might admit non-trivial hypercharge
flux. Furthermore, the existence of additional matter curves allows
one to realize a PQ-symmetry in a model with $4$--$1$ split, another
feature that was formerly ruled out, since it requires the existence
of at least three distinct $\rep{5}$ representations

It is interesting to note\footnote{We would like the thank Joe Marsano
  and Sakura Sch\"{a}fer-Nameki for pointing this out to us.} that the
$U(1)$-charges of $\Rep{5}$ and $\Rep{10}$ fields can be described by
a spectral cover with two $U(1)$ factors and ad-hoc breaking to a
certain diagonal $U(1)$. In fact, for any $3$--$1$--$1$ or
$2$--$2$--$1$ split one can define a $U(1)$ operator such that the
resulting $U(1)$ charges take the above form. For now, we simply note
that naive application of the Tate minimal model
algorithm~\cite{MR0393039, Katz:2011qp} does not yield a factorizing
spectral cover. A more detailed discussion of the non-uniqueness of
the Tate model will appear elsewhere~\cite{Braun:2013nqa}.

Finally, let us remark that we do not attempt to construct a
phenomenologically viable model in this paper. In particular, we will
not try to fix multiplicities to obtain three generations with the
correct doublet-triplet splitting or study the number of exotics in
the spectrum, as was done locally in~\cite{Marsano:2009gv,
  Marsano:2009wr, Dolan:2011iu, Marsano:2012yc}.  However, the point
that we do wish to make is that, in the context of F-theory
compactifications, there arise naturally Abelian factors in the gauge
group whose matter representations do not unify into a spontaneously
broken $E_8$. Such theories can easily circumvent phenomenological
stumbling blocks that have been found previously. A more complete
study of the systematics of $SU(5) \times U(1)$ models is
underway~\cite{Braun:2013nqa}.

\section{Calabi-Yau Geometry and its Sections} \label{s:cy}

In this section we introduce the fully resolved Calabi-Yau geometries
relevant for the $SU(5) \times U(1)$ GUT models with the spectrum
given in \autoref{s:field_theory}.  In \autoref{sec:elliptic}
we summarize some basics about elliptic fibrations. We argue that the
pattern of $U(1)$ charges naturally allows to introduce `splits'
independent of the factorization of defining equations or a spectral
cover. The Calabi-Yau threefold and fourfold examples supporting our
GUT spectrum are introduced in \autoref{sec:toric} and
\ref{ss:fourfold}. We carefully discuss the sections of these elliptic
fibrations.

\subsection{Basics on Elliptic Fibrations with \boldmath$SU(5)\times U(1)$ }
\label{sec:elliptic}

Before focusing on a specific example, let us quickly review the geometric
features of elliptic fibrations as relevant for F-theory models with
$SU(5) \times U(1)$ gauge symmetry. Compactification of F-theory
on an elliptically fibered threefold or fourfold yields a six- and 
four-dimensional effective theory respectively. 
In fact, one is interested in a singular Calabi-Yau manifold 
obtained by shrinking all irreducible fiber components not
intersecting the zero section. These singularities give rise to
non-Abelian gauge symmetries. However, arbitrary singularities are not
allowed and it is probably necessary that there exists a resolution to
a smooth Calabi-Yau. Therefore one can either consider smooth Calabi-Yau
manifolds where one can always contract the fiber components not
intersecting the zero section, or one can work with singular
Calabi-Yau manifolds but then has to prove that a resolution exists.

By definition, an elliptic fibration is a surjective map $\pi:X\to
\mathcal{B}$ from the Calabi-Yau variety to a base $\mathcal{B}$ whose
generic fiber is $T^2$ and such that there is a section. The fiber
degenerates along a divisor in the base $\mathcal{B}$ called the
discriminant. The generic degenerate fibers have been classified by
Kodaira and fall into an ADE-like pattern. For the purposes of this
paper, we will be mostly interested in the $I_n$ Kodaira fibers (the
$A_{n-1}$ series). Their dual graph (drawing a node for each $\CP^1$
irreducible component and a connecting line if they intersect) is the
cycle graph with $n$ nodes. Without monodromies the $I_n$ Kodaira fiber yields a
low-energy $A_{n-1}=SU(n)$ gauge theory, which is what we will be
interested in. More special fibers need not fall into the Kodaira
classification starting at codimension-two in the base (codimension
one along the discriminant). For starters, the codimension-two fibers
may actually be higher-dimensional, a feature that we like to avoid 
in our F-theory models.

We can distinguish two kinds of sections, and both will feature
prominently in this paper. The first and simpler case is that of a
holomorphic section (or just section), meaning that there is a holomorphic
embedding $s:\mathcal{B}\hookrightarrow X$ of the base in the elliptic
fibration such that the composition $\pi\circ s = \Id_\mathcal{B}$ is
the identity map on $B$. The second and more complicated case is that
of a rational section, that is, we require only a birational morphism
$s': \mathcal{B}\dashrightarrow \mathcal{B}' \subset X$ such that
$\pi\circ s' = \Id_\mathcal{B}$. This means that $s':\mathcal{B}\to
\mathcal{B}'$ is generically one-to-one, but not defined over some
points. The points where $s'$ cannot be defined is where the divisor
$\mathcal{B}'\subset X$ wraps a whole fiber component. Clearly, a
holomorphic section is a special case of a rational section, but we
stress that rational sections are perfectly fine for F-theory
compactifications. For physics applications, the rational sections give us
important additional freedom: A holomorphic section must intersect any
fiber in a single point, that is, it intersects a single irreducible
fiber component with intersection number one. Rational sections, on
the other hand, can wrap components of codimension-two fibers and
therefore have more freedom in the intersection numbers. This
translates into less constraints for the $U(1)$ matter charges, as we
will see in the following.

By definition, an elliptic fibration has at least one section. We pick
one and call it the zero section; physics does not depend on this
choice. The set of all sections then forms an Abelian group under
fiber-wise addition. That is, identify the torus $T^2 = \C / (\Z\oplus
\tau \Z)$ such that the zero section passes through $0\in \C$. Any
pair of section then passes through two points, which can be added in
$\C / (\Z\oplus \tau \Z)$ to obtain another section. The sections
together with this group law are called the \emph{Mordell-Weil} group
$MW(X)$. Its rank is the number of independent $U(1)$ gauge factors.
We will be mostly interested in the rank-one case in the remainder of
this paper, where there is an (up to sign) unique section that
generates $MW(X)\simeq \Z$.
\begin{figure}[htbp]
  \centering
  \includegraphics{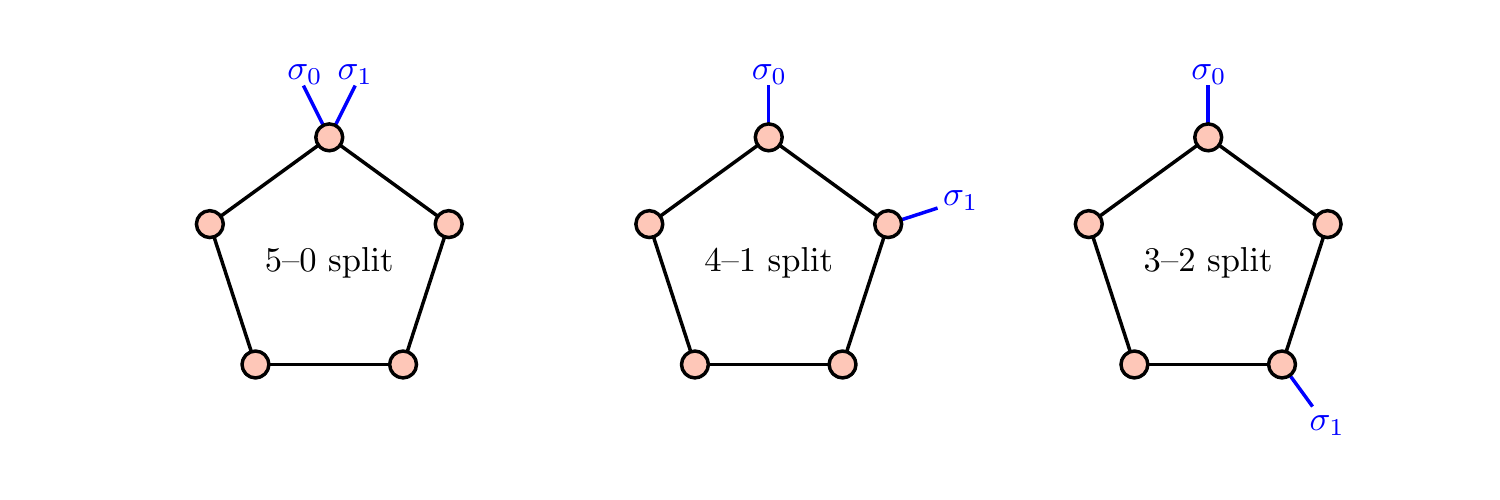}
  \vspace{-0.3cm}
  \caption{The three different relative orientations of the $I_5$
    discriminant and two sections $\sigma_0$, $\sigma_1$. This notion
    matches the splitting of the spectral cover if and only if the
    spectral cover unbroken gauge group is the low-energy effective
    $SU(5)\times U(1)$.}
  \label{fig:I5splits}
\end{figure}
To obtain an $SU(5)\times U(1)$ low-energy gauge theory, we thus need
an $I_5$ discriminant locus without monodromy together with two
sections (the zero section and the Mordell-Weil generator). Up to
relabelling, there are three different ways for the $I_5$ and two
sections to intersect which are depicted in \autoref{fig:I5splits}.

These three possibilities translate into different patterns of $U(1)$
charges for matter fields. Note that $U(1)$ charges can be
normalized arbitrarily, for example it is sometimes claimed that the
up quark has electric charge $\tfrac{2}{3}$. We will be using the sane
normalization where the minimal charge is one. With this nomenclature,
the three possibilities are
\begin{descriptionlist}
\item[$5$--$0$ split:] The $SU(5)$ singlets have minimal $U(1)$ charge
  one.
\item[$4$--$1$ split:] The $SU(5)$ singlets have $U(1)$ charges in
  $5\Z$. The $\Rep{5}$ of $SU(5)$ (fundamental representation) have
  $U(1)$ charge $2,3 \tmod 5$. The $\Rep{10}$ (antisymmetric
  representation) have $U(1)$ charges $1,4 \tmod 5$.
\item[$3$--$2$ split:] The $SU(5)$ singlets have $U(1)$ charges in
  $5\Z$. The fundamentals have $U(1)$ charges $1,4 \tmod 5$. The
  antisymmetrics have $U(1)$ charges $2,3 \tmod 5$.
\end{descriptionlist}
In special examples the two sections can be obtained by imposing that
some polynomial related to the Calabi-Yau hypersurface equation
factorizes. Note, however, that the $5$--$0$ split has obviously
nothing to do with a factorization of a degree-$5$ polynomial into a
degree-$5$ and a degree-$0$ polynomial and does therefore not appear
in the constructions of~\cite{Marsano:2009gv, Marsano:2009wr,
  Dudas:2010zb, Mayrhofer:2012zy}.

\subsection{Calabi-Yau Threefold and its Sections}
\label{sec:toric}

In this paper, we will be investigating F-theory compactifications on
elliptically fibered Calabi-Yau fourfolds. In particular, we will be interested
in the $U(1)$ charges of matter fields, which are determined by codimension-two fibers.
Hence this question is about the geometry of a
three-dimensional variety and nothing in the analysis changes in an
essential manner when we go from three- to fourfolds with the same fiber structure. Therefore, we 
will first discuss some crucial aspects in the context of Calabi-Yau 
threefolds in this section. For simplicity, we will be looking at
the simplest possible base $\CP^2$, though it would be very easy to
apply our methods to more complicated toric bases or to complete
intersections where the additional equations are constant along the
fiber direction.
\begin{table}
  \centering
  \begin{tabular}{rrrr|cc}
    \multicolumn{4}{c|}{Point $n_z \in \nabla \cap N$}
    & Coordinate $z$ & Divisor $V(z)$ \\
    \hline
    $-1$&$ -1$&$ -1$&$ -1$& $h_0$ &$ \Hhat{0}$ \\
    $ 0$&$  0$&$  0$&$  1$& $h_1$ &$ \Hhat{1}$ \\
    $-2$&$ -1$&$  1$&$  0$& $d_0$ &$ \Dhat{0}$ \\
    $-1$&$  0$&$  1$&$  0$& $d_1$ &$ \Dhat{1}$ \\
    $ 0$&$  0$&$  1$&$  0$& $d_2$ &$ \Dhat{2}$ \\
    $ 0$&$ -1$&$  1$&$  0$& $d_3$ &$ \Dhat{3}$ \\
    $-1$&$ -1$&$  1$&$  0$& $d_4$ &$ \Dhat{4}$ \\
    $-1$&$  0$&$  0$&$  0$& $f_0$ &$ \Fhat{0}$ \\
    $ 0$&$  1$&$  0$&$  0$& $f_1$ &$ \Fhat{1}$ \\
    $ 1$&$  0$&$  0$&$  0$& $f_2$ &$ \Fhat{2}$ \\
    $-1$&$ -1$&$  0$&$  0$& $f_3$ &$ \Fhat{3}$
  \end{tabular}
  \caption{The toric data for the smooth Calabi-Yau threefold $X$. Together
    with the origin, these are the only integral points in the
    lattice polytope $\nabla$ and we will be using the notation on the right
    for the corresponding toric divisors. The Hodge numbers are
    $h^{11}(X)=7$  and $h^{21}(X)=63$. Together with the fact that
    there is a $I_5$ discriminant component, the Shioda-Tate-Wazir
    formula~\cite{2001math.....12259W} tells us that $\rank MW(X)=1$.
    The fan is given in \eqref{fanSigma}.}
  \label{tab:X3}
\end{table}

To be completely explicit, we will be considering the Calabi-Yau
hypersurface~\cite{Grassi:2012qw} in the ambient toric variety
specified by \autoref{tab:X3}.  The hypersurface in the 4-d toric variety
is cut out by the equation
\begin{equation}
  \label{eq:hypersurf}
  p = \sum_{m \in \Delta\cap M} \alpha_m
  h_0^{n_{h_0}\cdot m + 1}
  h_1^{n_{h_1}\cdot m + 1}
  d_0^{n_{d_0}\cdot m + 1}
  \cdots
  f_3^{n_{f_3}\cdot m + 1}
  = 0
\end{equation}
where $\Delta$ is the dual (polar) polytope to $\nabla$, and $M$ is the dual lattice to $N$.  The elliptic
fibration is a toric morphism, that is, induced by a map of the fan
$\Sigma$ of the toric ambient space, given explicitly in
Eq.~\eqref{fanSigma}, to the fan of $\CP^2$ by projecting on the last
two coordinates of $N\simeq \Z^4$. In terms of homogeneous
coordinates, the projection map $\pi: X \to \CP^2$ is given by
\begin{equation}
  \label{eq:fib}
  \pi:\
  [h_0:h_1:d_0:\ldots:d_4:f_0:\ldots:f_3] 
    \mapsto
  [h_0:h_1: d_0 d_1 d_2 d_3 d_4]
\end{equation}
We see that the homogeneous coordinates $f_0$, $\dots$, $f_3$
corresponding to the rays in the kernel of the projection parametrize
the fiber in the ambient space. The $I_5$ discriminant component is
the curve $[z_0:z_1:0]\in \CP^2$ and the $5$ divisors $\Dhat{0}$,
$\dots$, $\Dhat{4}$ map to it. In a generic fiber of the discriminant
(codimension-one over the base), the Calabi-Yau hypersurface cuts out
a $\CP^1$ in each of the 5 components, yielding the $I_5$ Kodaira
fiber.

The simplest way to define a section is to pick a toric divisor on the
generic fiber, that is, set one of the $f_i$ to zero. Together with
the hypersurface equation, this cuts out a certain number of points in
each fiber. The number of points can be computed using intersection
theory, or naively by plugging in $f_i=0$ into the hypersurface
equation. Homogeneous coordinates whose points are not in the star
of the cone $\langle n_{f_0}\rangle$ cannot vanish simultaneously with
$f_0$ and can be scaled to one.\footnote{These coordinates lie in the Stanley-Reisner
ideal when multiplied with $f_0$} 
Setting $f_0=0$, $f_2=d_i=1, i>0$ the 
hypersurface equation \eqref{eq:hypersurf} takes the form 
\begin{equation}
  p:\quad 
  \alpha_0 f_1 +
  \big(
  \alpha_1 h_0^2 + \alpha_2 h_0 h_1 + \alpha_3 h_1^2 + 
  \alpha_4 h_0 d_0  + \alpha_5 h_1 d_0 + \alpha_6 d_0^2
  \big) f_3
  = 0
  .
\end{equation}
This equation can be solved trivially for the homogeneous fiber
coordinates $[f_1:f_3]$ along the $\Fhat{0}$ divisor. In fact,
$f_1\not=0=f_3$ is forbidden if all coefficients $\alpha_m$ are
sufficiently generic, so we may scale $f_3=1$ as well. Thus, the
section is
\begin{equation}
  \label{eq:sigma0}
  \begin{gathered}
    \sigma_0: \quad [h_0:h_1:d_0] \mapsto 
    [h_0: h_1: d_0: 1: 1: 1: 1: 0: f_1(h_0,h_1,f_0) : 1: 1]
    ,\\
    f_1(h_0,h_1,f_0) =
    -\tfrac{1}{\alpha_0}
    \big(
    \alpha_1 h_0^2 + \alpha_2 h_0 h_1 + \alpha_3 h_1^2 + 
    \alpha_4 h_0 d_0  + \alpha_5 h_1 d_0 + \alpha_6 d_0^2
    \big)
    .
  \end{gathered}
\end{equation}
We see that $\sigma_0 = \{ p = f_0=0 \}$ is not only a section, which
could have been learned from intersection theory alone, but also that
it is a holomorphic section.


It remains to find a second section, namely the generator of the
Mordell-Weil group. This is made more interesting by the fact that
none of the remaining toric fiber divisors $\Fhat{1}$, $\Fhat{2}$,
$\Fhat{3}$ defines a section for us. In fact, $\Fhat{1}$ and
$\Fhat{3}$ define two-sections and $\Fhat{2}$ a three-section. Hence
we will approach this section differently, and, instead of explicitly
finding its equation, we will determine its homology class. A first
guess, which is wrong but instructive, is to take
$[\Fhat{1}-\Fhat{0}]$. It is a two-section minus a section and therefore,
numerically, a section. In more elaborate terms,\footnote{Note that
  the divisors $\Hhat{0}=\pi^{-1}([0:*:*])$ and
  $\Hhat{1}=\pi^{-1}([*:0:*])$ are elliptic fibrations over the
  coordinate $\CP^1$ in the base that intersect the discriminant
  transversely.} the generic fiber has the homology class
$\Hhat{0}\cap \Hhat{1} = \pi^{-1}([0:0:1])$. By a simple intersection
computation, its intersection with the tentative section is therefore
\begin{equation}
  [\Fhat{1}-\Fhat{0}] \cap \Hhat{0}\cap \Hhat{1} = 1
  .
\end{equation}
However, other intersection numbers show that the class
$[\Fhat{1}-\Fhat{0}]$ does not contain a section. By intersecting the
fibral\footnote{The \emph{fibral} divisors $\Dhat{i}$ are the divisors
  swept out by irreducible components of the $I_5$ Kodaira fiber as we
  move the curves along over the discriminant.} divisors with
$\Hhat{0}$, $\Hhat{1}$ we obtain the irreducible component curves
$C_i\simeq \CP^1$ of the $I_5$ Kodaira fibers as
\begin{equation}
  C_i = \Dhat{i} \cap \Hhat{0} = \Dhat{i}\cap \Hhat{1}
  .
\end{equation}
Computing the intersection numbers with the tentative section, we
obtain
\begin{equation}
  [\Fhat{1}-\Fhat{0}] \cap C_i = 
  [\Fhat{1}-\Fhat{0}] \cap \Dhat{i} \cap \Hhat{0} 
  =
  \begin{cases}
    -1 & i=0, \\
    1 & i=1,2, \\
    0 & i=3,4.
  \end{cases}
\end{equation}
The fact that the intersection number is negative means that the $I_5$
component curve $C_0$ is contained $[\Fhat{1}-\Fhat{0}]$ as we slide
it along over the discriminant. That is, the whole fibral divisor
$\Dhat{0}$ is contained in $[\Fhat{1}-\Fhat{0}]$. But since a rational
section may only contain components of codimension-two fibers and not
complete fibral divisors (which are codimension-one over the
base), $[\Fhat{1}-\Fhat{0}]$ is not a rational section after
all. However, it is clear that this can be fixed by subtracting the fibral divisor
$\Dhat{0}$. 

Therefore our new best guess for the class of the section generating the
Mordell-Weil group is $[\Fhat{1}-\Fhat{0}-\Dhat{0}]$. Computing
intersection numbers, one finds that it still does not work and
one needs to subtract further vertical divisors. After repeating the same steps several times, the end result is the
homology class
\begin{equation} \label{eq:MWsection}
  [\sigma_1]= 
  [\Fhat{1}-\Fhat{0}-\Dhat{0}-\Dhat{3}-\Dhat{4}+\Hhat{0}].
\end{equation}
Showing that this homology class actually contains a section is more
involved and will be presented in \autoref{app:showsection}.  This is
the section generating the Mordell-Weil group and, as we will see in
the following, it is only a rational section. Computing the
intersection number $\sigma_1 \cap \Dhat{1}\cap \Hhat{0}$ and noticing
that $\sigma_0$ intersects $\Dhat{0}$ only from Eq.~\eqref{eq:sigma0},
we note that this elliptic fibraiton is of the $4$--$1$ split
type. Finally, we note from the sheaf cohomology computation that the
section $\sigma_1$ exists only on the Calabi-Yau hypersurface and does
not extend to a section on the whole ambient toric variety. This is
why its construction has been so tedious.

\subsection{Calabi-Yau Fourfold and its Sections} \label{ss:fourfold}

As we have mentioned previously, the matter content induced by a
particular Calabi-Yau manifold depends on the codimension-two
singularities of the variety and one can extend the base from
$\mathbb{P}^2$ to $\mathbb{P}^3$ without changing the types of matter
representations that can occur. Extending the toric data in
\autoref{tab:X3} to a fourfold is therefore very simple: We embed the
points $n_z$ into $N' \simeq \mathbb{Z}^5$ via
\begin{equation}
  n_z \mapsto \left\{
    \begin{array}{ll}
      (n_z,0) & \ \textrm{for} \ z \neq h_0 \\
      (n_z,-1) & \ \textrm{for} \ z = h_0 \\
    \end{array}
  \right.
  \label{eq:X4rays}
\end{equation}
and add an additional point $n_{h_2} = (0,0,0,0,1)$ whose divisor we
denote by $\hat{H}_2$. By abuse of notation, we will use the same
letters for divisors and homogeneous coordinates. Since we will never
be talking about the Calabi-Yau fourfold and the threefold at the
same time, their meaning should be clear from context.

Of the six triangulations that induce inequivalent intersection
numbers on the Calabi-Yau hypersurface in the ambient toric fivefold,
we choose the one given by \eqref{fanSigma2}. We remark that for this
choice the zero section corresponding to $f_0 = 0$ is not holomorphic
anymore. However, this no problem for F-theory appliations. By the
exact same reasoning as before, we find that there is (up to sign) a
single Mordell-Weil generator. Its homology class is almost
identical to the one in Eq. \eqref{eq:MWsection}, namely
\begin{equation} \label{eq:MWSection_4d}
  [\sigma_1] = [\Fhat{1}-\Fhat{0}-\Dhat{0}-\Dhat{3}-\Dhat{4}+2 \Hhat{0}]\;.
\end{equation}  
To make contact with the notation in~\cite{Cvetic:2012xn,
  Grimm:2011sk, Grimm:2010ks} let us define a shifted bases of
divisors, which we denote by unhatted letters. Most of the
redefinitions are trivial, namely
\begin{equation}
  D_i = \Dhat{i}  \quad \text{for} \ i \neq 0,\quad 
  F_i = \Fhat{i}, \quad \text{and} \quad
  H_i = \Hhat{i} ,
\end{equation}
but the following have a deeper meaning:
\begin{equation}
  \begin{split}
    D_0 &= \Fhat{0} + 2 \Hhat{1} \\
    D_5 &= -39 \Dhat{2} - 76 \Dhat{3} - 38 \Dhat{4} + 7 \Fhat{3} + 39
    \Fhat{1} - 41 \Fhat{2} + 31 \Fhat{0}
  \end{split}
\end{equation}
$D_0$ represents a physically motivated choice for our base divisor in
terms of which the low-energy effective action takes a more convenient
form, while $D_5$ is the Abelian $U(1)$-divisor associated with the
Mordell-Weil generator in the class $[\sigma_1]$ and is obtained via
the Shioda map. A convenient basis of independent divisors is given by
the set
\begin{align}
  (D_0, D_\alpha, D_i, D_m) \equiv (D_0, H_1, D_i, D_5)\;.
\end{align}
Here $D_\alpha, \ \alpha=1,\ldots,h^{1,1}(\mathcal{B}_3)$ are
so-called vertical divisors defined as pre-images of base divisors
under the projection, i.e. $D_\alpha = \pi^{-1}(D_\alpha^b)$. In this
particular case $h^{1,1}(\mathbb{P}^3) =1$ and there is only a single
independent vertical divisor.

Having introduced this notation, we can compute the Green-Schwarz coefficients appearing in Eq. \eqref{e:su5_anomalies}.
$a^\alpha$ is given by
\begin{align}
  c_1(\mathcal{B}_3) = a^\alpha D_\alpha^b\;,
\end{align}
and therefore $a^\alpha = 2$ for our model, while $b^\alpha$ and $b^\alpha_{(5)}$ are the expansion coefficients
of the base divisors wrapped by the GUT-divisor and the brane supporting the $U(1)$-factor, respectively.
Here, they can be computed to be
\begin{align}
  b^\alpha = - \frac{D_i^2 \cdot H^2}{B \cdot H^3} = 1 \quad \textrm{and} \quad b^\alpha_{(5)} =
  - \frac{D_5^2 \cdot H^2}{B \cdot H^3} = 570
\end{align}
and we refer to~\cite{Cvetic:2012xn} for the general case.

Furthermore, let us comment on the fact that the appearance of a
non-holomorphic zero section means that not all of the intersection
numbers given in~\cite{Cvetic:2012xn} hold anymore. To be precise, one
only has
\begin{align}
  D_0 \cdot D_\Lambda \cdot D_\alpha \cdot D_\beta = 0\;
\end{align}
instead of $[D_0 \cdot D_\Lambda] = 0$ in the cohomology of all of $X_4$, as holds for holomorphic zero sections.

Finally, we would like to remark that unlike the Calabi-Yau threefold example,
the fourfold presented here possesses a non-flat fibration, that is, there exist isolated points
in the base over which the fiber component $C_4$ becomes two 
complex-dimensional.~\footnote{We thank Andreas Braun, Christoph Mayrhofer, and Sakura Sch\"{a}fer-Nameki for
discussions on this point.}
As noted earlier, this particular feature is not desirable for phenomenological reasons. However, this problem
can easily be remedied by substituting the base manifold $\mathbb{P}^3$ with a different one
such as $\mathbb{P}^2 \times \mathbb{P}^1$. In particular, this substitution has no
impact on the non-Abelian matter content and its $U(1)$-charges. For the purpose
of the discussions in this paper, we can therefore safely ignore this issue and 
refer to \cite{Braun:2013nqa} for a more detailed discussion of flatness and the flat example with a
$\mathbb{P}^2 \times \mathbb{P}^1$ base.

\section{Matter and Abelian Charge Assignments}
\label{sec:charge}

In this section we derive the matter spectrum present in the
Calabi-Yau threefold and Calabi-Yau fourfold
compactifications. Focusing first on the Calabi-Yau threefold we
discuss the intersection theory and the split of the resolving fibers
in \autoref{sec:intersection}.  The $U(1)$ charges of the fundamental
matter are obtained in \autoref{sec:fundamental}.  We comment on the
lift of these results to the Calabi-Yau fourfold. The four-dimensional
chiral spectrum is then computed in \autoref{s:chiralities}.

\subsection{Intersection Theory}
\label{sec:intersection}

By computing the discriminant of the elliptic fibration as a
degree-$36$ polynomial over the base $\CP^2$
explicitly~\cite{Braun:2011ux}, one can always enumerate the
codimension-two fibers where the $I_5$ Kodaira fiber degenerates
further. We now pick a sufficiently generic hypersurface using random
coefficients in Eq.~\eqref{eq:hypersurf}, find the location
of the codimension-two fibers numerically, and analyze the
hypersurface in these special fibers. Roughly, the hypersurface will
factorize in one of the irreducible components of the toric ambient
fiber, and this defines the charge of the localized matter field.

Naively, we face an impasse: the combinatorial description of the
geometry of the ambient toric variety knows nothing about whether a
hypersurface equation factorizes or not. Hence no toric intersection
computation on the toric variety $X_\Sigma$ can possibly capture the
irreducible curves that are stuck on the codimension-two fiber; but
the zero modes on those curves are precisely the matter fields that we
are after. However, this argument is a bit too simple minded and,
while we cannot use simply intersection theory on $X_\Sigma$, toric
methods still apply. The trick is to construct the irreducible
components of the fibers of the ambient space, which are
two-dimensional toric varieties. The hypersurface restricted to the
ambient toric fiber will factorize into multiple irreducible
components, each of which has its own divisor class on the
surface. Then all that remains is to pull back the sections to this
fiber component and apply the usual toric intersection theory there.

To clarify this procedure, let us look at an example and consider the
irreducible fiber component $C_0 = \Dhat{0} \cap \Hhat{0}$ of the
$I_5$ Kodaira fiber that intersects the zero-section $\sigma_0$. The
star of the corresponding ray $\langle n_{d_0} \rangle$ contains the
homogeneous coordinates $h_0$, $h_1$, $d_1$, $d_2$, $d_4$, $f_0$, $f_1$, and
$f_3$. We set $d_0$ to zero and all remaining variables to
one. According to the fibration map Eq.~\eqref{eq:fib}, the point on
the $I_5$ discriminant locus $[h_0:h_1:0]\in \CP^2$ is parametrized by
the ratio of $h_0$ and $h_1$, which we treat in the following as
numerical constants that have been fixed to restrict us to a
particular codimension-two fiber. Plugging this into the hypersurface
equation, we obtain four non-zero terms
\begin{equation}
  \label{eq:hyperD0}
  p(h_0, h_1, 0, d_1, d_2, 1, d_4, f_0, f_1, 1, f_3) = 
  \beta_0 d_1 d_2^2 d_4 f_1 + \beta_1 d_1 d_2 f_0 f_1^2 + 
  \beta_2 d_2 d_4 f_3 + \beta_3 f_0 f_1 f_3
\end{equation}
where $\beta_0$, $\dots$, $\beta_3$ are constants depending on the
fixed $h_0$, $h_1$. 

For special values of the $h_0$, $h_1$ the coefficients $\beta_i$
become special and the hypersurface equation factorizes. This is how
the $I_5$ Kodaira fiber degenerates further at codimension-two
fibers. A computation shows that~\cite{Morrison:2011mb}
\begin{itemize}
\item at $2$ distinct codimension-two fibers the coefficient $\beta_2$
  vanishes and the polynomial factorizes as
  \begin{equation}
    \label{eq:factor2}
    p(h_0, h_1, 0, d_1, d_2, 1, d_4, f_0, f_1, 1, f_3) = 
    f_1 \times
    \big( \beta_0 d_1 d_2^2 d_4 + \beta_1 d_1 d_2 f_0 f_1 + 
     \beta_3 f_0 f_3 \big)
  \end{equation}
\item at $3$ distinct codimension-two fibers the hypersurface
  equation factors as
  \begin{equation}
    \label{eq:factor3}
    p(h_0, h_1, 0, d_1, d_2, 1, d_4, f_0, f_1, 1, f_3) = 
    (\beta_0' d_1 d_2 f_1 + \beta_1' f_3) \times    
    (\beta_2' d_2 d_4 + \beta_3' f_0 f_1) 
  \end{equation}
\item at further $14$ codimension-two degenerate fibers the
  hypersurface equation on the fiber component $C_0$ does not
  factorize. Instead, other irreducible components of the $I_5$ fiber,
  that is, $C_i =\Dhat{i} \cap \Hhat{0}$ for $i\not=0$, become
  reducible.
\item finally, there are $3$ remaining codimenion-two fibers where
  multiple $I_5$ components factor simultaneously. This is where the
  $\Rep{10}$ matter fields are localized.
\end{itemize}

To understand the intersection theory on the fiber, we have to
construct the fiber component $C_0 = \Dhat{0} \cap \Hhat{0}$ as a
toric variety. That is, the remaining homogeneous coordinates $d_1$,
$d_2$, $d_4$, $f_0$, $f_1$, $f_3$ on the right hand side of
Eq.~\eqref{eq:hyperD0} are the homogeneous coordinates of a
two-dimensional toric variety. The toric surface can be reconstructed
from knowing how the homogeneous coordinate rescalings act. First, one
has to identify the subset of homogeneous rescalings on the 4-d toric
variety $X_\Sigma$ that do not change the values of $h_0$ and
$h_1$. Then, ignore the action on $d_0$ since it is being set to
\begin{figure}[htbp]
  \centering
  \begin{minipage}{0.5\linewidth}
    \includegraphics[width=5.5cm]{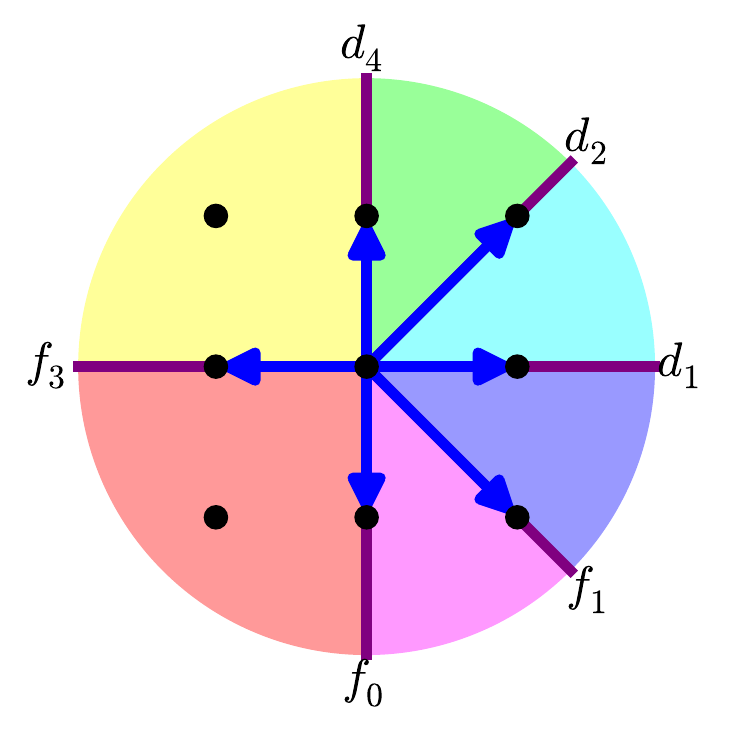}
  \end{minipage}
  \begin{minipage}{0.35\linewidth}
      \begin{tabular}{rr|c|c}
        \multicolumn{2}{c|}{Point $n_z$}
        & Coord. $z$ & $V(z)$ \\
        \hline
        $ 1$&$  0$& $d_1$ &$ \cDbar{1}$ \\
        $ 1$&$  1$& $d_2$ &$ \cDbar{2}$ \\
        $ 0$&$  1$& $d_4$ &$ \cDbar{4}$ \\
        $-1$&$  0$& $f_3$ &$ \cFbar{3}$ \\
        $ 1$&$ -1$& $f_1$ &$ \cFbar{1}$ \\
        $ 0$&$ -1$& $f_0$ &$ \cFbar{0}$ \\
      \end{tabular}
  \end{minipage}
  \caption{The toric ambient space fiber $C_0$, that is, one of the
   five irreducible components of $\pi^{-1}([h_0:h_1:0])$.}
  \label{fig:C0fan}
\end{figure}
zero. The result is that the toric surface on which Eq.~\eqref{eq:hyperD0}
is defined is the one shown in \autoref{fig:C0fan}. In more
elaborate terms, this is the relative star construction
of~\cite{2000math.....10082H}. This toric surface is embedded into the
fiber of the toric variety $X_\sigma$ over $[h_0:h_1:0]$ via
\begin{equation}
  \label{eq:i0}
  i_0: [d_1:d_2:d_4:f_0:f_1:f_3] 
  \mapsto
  [h_0: h_1: 0: d_1: d_2: 1: d_4: f_0: f_1: 1: f_3] 
\end{equation}

We now take advantage of the toric surface description of the fiber
component. First, we can formulate the factorization of the
hypersurface equation as follows:
\begin{itemize}
\item At $2$ distinct codimension-two fibers, where the hypersurface
  factors as in Eq.~\eqref{eq:factor2}, the $I_5$ fiber component
  splits into two irreducible components with homology classes
  \begin{equation}
    \label{eq:factor2split_1}
    V(p) = \big( \cFbar{1} \big ) + 
    \big( \cFbar{0} + \cFbar{3} \big),
  \end{equation}
\item and at $3$ distinct codimension-two fibers, where the
  hypersurface equation factors as Eq.~\eqref{eq:factor3}, the $I_5$
  fiber component splits into two irreducible components with homology
  classes 
  \begin{equation}
    \label{eq:factor2split_2}
    V(p) = \big(\cFbar{0} + \cFbar{1}\big) + 
    \big( \cFbar{3}\big).
  \end{equation}
\end{itemize}
Furthermore, the sections $\sigma_0$, $\sigma_1$, as divisors on
$X_\Sigma$, can be pulled back by the embedding map $i_0$, see
Eq.~\eqref{eq:i0}. The details of the toric algorithm for the pullback
by the fiber embedding can be found in~\cite{2000math.....10082H}. The
result is that
\begin{equation}
  \begin{split}
    i_0^*(\sigma_0) =&
    \cFbar{0}
    , \\
    i_0^*(\sigma_1) =&
    \cFbar{3} -  \cFbar{0}
    .
  \end{split}
\end{equation}
To summarize, the $I_5$ Kodaira fiber degenerates at $2+3$
codimension-two fibers by splitting the irreducible component
intersecting the zero-section in two, yielding a fiber of Kodaira type
$I_6$. However, in the first two fibers it splits into two curves that
are distinct from the split in the last $3$ fibers. The fiber
components and their intersection number with the sections is given in
\autoref{tab:intersection}.
\begin{table}[htbp]
  \centering
  \begin{tabular}{c|cccccc}
    $I_6$ component &
    $\bar{C}_0$ & 
    $\bar{C}_1$ & 
    $\bar{C}_2$ & 
    $\bar{C}_3$ & 
    $\bar{C}_4$ & 
    $\bar{C}_5$ \\
    \hline
    Realization &
    $\cFbar{0} + \cFbar{1}$ & 
    $\cFbar{3}$ & 
    $C_1$ & 
    $C_2$ & 
    $C_3$ & 
    $C_4$ \\
    $\cap \sigma_0$ & 
    $ 0 $&$ 1 $&$ 0 $&$ 0 $&$ 0 $&$ 0 $ \\
    $\cap \sigma_1$ & 
    $ 1 $&$ -1 $&$ 0 $&$ 0 $&$ 1 $&$ 0 $
  \end{tabular}
  \vspace{5mm}

  \begin{tabular}{c|cccccc}
    $I_6$ component &
    $\bar{C}_0$ & 
    $\bar{C}_1$ & 
    $\bar{C}_2$ & 
    $\bar{C}_3$ & 
    $\bar{C}_4$ & 
    $\bar{C}_5$ \\
    \hline
    Realization &
    $\cFbar{3}$ &
    $\cFbar{0} + \cFbar{1}$ &
    $C_1$ & 
    $C_2$ & 
    $C_3$ & 
    $C_4$ \\
    $\cap \sigma_0$ & 
    $ 1 $&$ 0 $&$ 0 $&$ 0 $&$ 0 $&$ 0 $ \\
    $\cap \sigma_1$ & 
    $ -1 $&$ 1 $&$ 0 $&$ 0 $&$ 1 $&$ 0 $
  \end{tabular}
  \caption{Intersection numbers of the two different $I_6$-type
    codimension-two fibers where the codimension-one $I_5$ fiber
    splits the fiber component intersecting the zero section. The
    curves $\bar{C}_i$ are the $I_6$ fiber components in cyclic
    order. The curves $C_i$ are the $I_5$ fiber components $C_i =
    \Dhat{i} \cap \Hhat{0}$.}
  \label{tab:intersection}
\end{table}

\subsection{Fundamental Matter}
\label{sec:fundamental}

The two different degenerations of the $I_5$ Kodaira fiber into
codimension-two $I_6$-type fibers result in localized $2\times\Rep{5}$
and $3\times\Rep{5}$ matter of $SU(5)$. They will turn out to be
distinguished by their $U(1)$ charge, as we are about to see. The
$U(1)$ charge is given by the intersection of the curves stuck at
codimension-two fiber, that is, the irreducible components of the
factored $I_5$ component, with the image of the section under the
Shioda map~\cite{Morrison:2012ei} $S:MW(X)\to H_4(X,\Q)$. For a single
$I_5$ Kodaira fiber, this boils down to
\begin{equation}
  \begin{split}
    U(1)\text{-charge}(\bar{C}_i) &=
    \bar{C}_i \cap S(\sigma_1)
    \\[-2ex] &=
    \bar{C}_i \cap \sigma_1 - \bar{C}_i \cap \sigma_0 
    + 
    \sum_{1\leq a,b \leq 4}
    (\bar{C}_i \cap \Dhat{a})
    \left(\begin{smallmatrix}
        \frac{4}{5} & \frac{3}{5} & \frac{2}{5} & \frac{1}{5} \\
        \frac{3}{5} & \frac{6}{5} & \frac{4}{5} & \frac{2}{5} \\
        \frac{2}{5} & \frac{4}{5} & \frac{6}{5} & \frac{3}{5} \\
        \frac{1}{5} & \frac{2}{5} & \frac{3}{5} & \frac{4}{5}
      \end{smallmatrix}\right)_{ab}
    (\sigma_1 \cap C_{b})
  \end{split}
\end{equation}
For example, consider $\bar{C}_0 = \cFbar{0} + \cFbar{1}$, a curve
contributing to the $2\times\Rep{5}$. Its intersections with
$\sigma_0$, $\sigma_1$ are listed in the upper half of
\autoref{tab:intersection}.
\begin{equation}
  U(1)-\text{charge}(2\times \Rep{5}) = 
  1 - 0 + 
  \left(\begin{smallmatrix}
      0 & 0 & 0 & 1
    \end{smallmatrix}\right)
  \left(\begin{smallmatrix}
      \frac{4}{5} & \frac{3}{5} & \frac{2}{5} & \frac{1}{5} \\
      \frac{3}{5} & \frac{6}{5} & \frac{4}{5} & \frac{2}{5} \\
      \frac{2}{5} & \frac{4}{5} & \frac{6}{5} & \frac{3}{5} \\
      \frac{1}{5} & \frac{2}{5} & \frac{3}{5} & \frac{4}{5}
    \end{smallmatrix}\right)
  \left(\begin{smallmatrix}
      0 \\ 0 \\ 1 \\ 0
    \end{smallmatrix}\right)
  =
  \frac{8}{5}
\end{equation}
Similarly, the $U(1)$ charge of the other $3\times \Rep{5}$ ends up
being $\tfrac{7}{5}$. As noted in \autoref{sec:intersection}, there
are further $14$ codimension-two fibers giving rise to $\Rep{5}$ and 3
more yielding $\Rep{10}$ matter. Their $U(1)$ charge can be computed
by straightforward application of the same methods and we will leave
the details as an exercise to the reader. The result is that, after
clearing denominators to make the $U(1)$ charges integral, the
$SU(5)$-charged spectrum is
\begin{equation}
  2\times \Rep{5}_8 + 
  3\times \Rep{5}_7 +
  6\times \Rep{5}_3 +
  8\times \Rep{5}_2 +
  3\times \Rep{10}_1
  .
\end{equation}
The Calabi-Yau fourfold will have the same types of representations
arising, since they are determined by the behaviour at a generic point
on a matter curve. In other words, after intersecting the matter curve
with a divisor crossing it, the same analysis for the $SU(5)\times
U(1)$ representation content applies. Of course, the 6-d quaternionic
representations will be split up into conjugate pairs of 4-d
representations, and the multiplicity of the representations will be
different. In fact, the multiplicities do depend on the four-form flux
which is a phenomenon for fourfolds that has no threefold analogue,
and will be the topic of the following section.

\section{Chiral Index from One-Loop Chern-Simons Terms}
\label{s:chiralities}

In this final section we compute the four-dimensional chiral indexes
for the matter spectrum induced by an F-theory compactification on the
Calabi-Yau fourfold of \autoref{ss:fourfold}.  We first make some
general remarks on $G_4$ fluxes and their induced three-dimensional
Chern-Simons terms in \autoref{sec:G4generalities}. The explicit
computations of the chiral indices for our example are presented in
\autoref{sec:chiralspec}.

\subsection{General Remarks on \boldmath $G_4$-Fluxes and
  Chiralities} \label{sec:G4generalities}
Naively reducing the six-dimensional matter multiplets to four dimensions, the resulting matter representations appear in vector-like pairs. Thus, the resulting theory is non-chiral. However, this changes as soon as fourform fluxes are included on the M-theory side. Effectively, their inclusion is equivalent to projecting out certain matter multiplets, hence leading to non-trivial chiral indices. The complete data of an F-theory compactification to four dimensions therefore consists of the Calabi-Yau fourfold $X_4$ supplemented by a choice of G-flux satisfying certain consistency conditions. We will now explain what these conditions are, construct a concrete set of fluxes and calculate the resulting chiral indices.

First of all, a consistent choice of $G_4$ must satisfy a quantisation condition \cite{Witten:1996md}, i.e.
\begin{align}
 G_4 + \tfrac{1}{2} c_2(\hat X_4) \in H^4(\hat{X}_4, \mathbb{Z})\ ,
\end{align}
where $c_2(\hat X_4)$ is the second Chern class of $\hat X_4$.
Furthermore, the fourth cohomology group of $\hat{X}_4$ splits into orthogonal subspaces, namely a horizontal one whose elements are derived from complex structure variations of the global $(4,0)$-form on $\hat{X}_4$ and a vertical one. The vertical subspace contains four-forms obtained by taking the wedge product of two $(1,1)$-forms on $\hat{X}_4$. While horizontal fluxes give rise to a non-trivial flux superpotential, the vertical fluxes induce chirality in the four-dimensional theory and we therefore concentrate on them. For more details
on flux quantisation in F-theory, we refer to \cite{Collinucci:2010gz,Collinucci:2012as}. 
Chirality induced by G-fluxes has recently been studied in \cite{Donagi:2008ca,Beasley:2008dc, Beasley:2008kw,Hayashi:2008ba,Marsano:2011hv,Braun:2011zm,Krause:2011xj,Grimm:2011fx,Kuntzler:2012bu,Cvetic:2012xn}. We will follow the general approach of \cite{Grimm:2011fx,Cvetic:2012xn}.

In order to determine the chiral indices a given flux induces, we employ a result obtained by using M-/F-theory duality in three and four dimensions, respectively. On the M-theory side, non-trivial G-flux induces a Chern-Simons term in three dimensions
\begin{align}
 S_{CS}^{(3)} = -\frac{1}{2} \int \Theta_{A B}^M A^A \wedge F^B\;,
\end{align}
with flux coefficients
\begin{align}
 \Theta_{A B}^M = \frac{1}{2} \int_{X_4} G_4 \wedge \omega_A \wedge \omega_B\;.
\end{align}
and $\omega_{A}$ the basis of $(1,1)$-forms introduced in
\autoref{ss:fourfold}. On the F-theory side, Chern-Simons terms
originate from integrating out charged matter in the circle reduction
from four to three dimensions and one can show that their coefficients
are given by
\begin{align}
 \Theta^F_{\Lambda \Sigma} = - \sum_{\rep{R}} \chi(\rep{R}) \sum_{\mathbf{w} \in \rep{R}} \mathbf{w_\Lambda} \mathbf{w_\Sigma} \sign (\mathbf{w}) \label{e:CS_exceptional}
\end{align}
Here $\Lambda, \Sigma = (i,5)$, $\mathbf{R}$ runs over all complex
representations and $\mathbf{w}$ are the weights of a representation
$\rep{R}$. To each weight $\mathbf{w}$ one can naturally assign a
curve in $X_4$ and we take $\sign(\mathbf{w})$ to be $+1$ if this
curves shrinks to a point in the F-theory limit $X_4 \to \hat{X_4}$
and $-1$ otherwise. For details of how to calculate
$\sign(\mathbf{w})$ we refer to~\cite{Grimm:2011fx, Cvetic:2012xn},
where analogous computations were carried out.

Before giving an explicit form of $G_4$, we impose some additional constraints:
\begin{equation}
  \begin{gathered}
    \Theta_{i \alpha} = \Theta_{\alpha \beta} = \Theta_{0 \alpha} = 0 \\
    \Theta_{0 i} = \Theta_{0 0} = 0 \label{e:flux_constraints}
  \end{gathered}
\end{equation}
The first three equations forbid a broken non-Abelian gauge group, non-geometric fluxes and fluxes along the 3d/4d-circle respectively. For holomorphic zero sections the remaining two constraints follow automatically from the first three and therefore we impose them by hand for our rational zero section.

After enforcing the constraints in Eq.~\eqref{e:flux_constraints}, the only other non-vanishing Chern-Simons terms apart from \eqref{e:CS_exceptional} are $\Theta_{05}$. A loop calculation on the F-Theory side determines them to be \cite{Cvetic:2012xn}
\begin{align}
 \Theta^F_{05} = \frac{1}{6} \sum_q n(q) q\;,
\end{align}
where $n(q)$ counts the number of fields with $U(1)$ charge q. For example, in the case of our spectrum one has that 
\begin{align}
 n(-1) - n(+1) = 10 \chi(\rep{10_{-1}})
\end{align}
since every $\rep{10}$ representation has ten different weights. Note that the matching condition $\Theta^F_{05} = \Theta^M_{05}$ is precisely equivalent to the cancellation of the gravitational-Abelian anomaly in Eq. \eqref{e:su5_anomalies}, since $\Theta_{05} = \frac{1}{2} \Theta_{\alpha 5}$.

\subsection{Chiral Matter Spectrum in the Example} \label{sec:chiralspec}

For the triangulation \eqref{fanSigma2} we find four independent flux parameters, namely
\begin{equation}
  \begin{split}
    G_4 =&\, \alpha \Big( 2 D_{2}^{2} - 28 D_{3} D_{4} - 7 D_{4}^{2} + \tfrac{9}{4} F_{3}^{2} -
    \tfrac{28}{5} D_{2} F_{1} + 4 F_{1}^{2} + \tfrac{11}{10} F_{3} F_{2} +
    \tfrac{122}{15} F_{1} F_{2} - \tfrac{36}{5} F_{2}^{2}   \\
    &- \tfrac{65}{6} F_{1}
    F_{0} - 9 F_{0}^{2} \Big)  + \beta \Big( - D_{2}^{2} + 8 D_{3} D_{4} + 2 D_{4}^{2} - \tfrac{1}{2} F_{3}^{2} +
    \tfrac{6}{5} D_{2} F_{1} - \tfrac{1}{5} F_{3} F_{2} - \tfrac{49}{15} F_{1} F_{2}  \\ 
    &+ \tfrac{12}{5} F_{2}^{2} + 3 F_{1} F_{0} + 2 F_{0}^{2} \Big) 
    + \gamma \Big( - D_{2}^{2} + 8 D_{3} D_{4} + 2 D_{4}^{2} - \tfrac{1}{2} F_{3}^{2} +
    \tfrac{6}{5} D_{2} F_{1} + \tfrac{11}{5} F_{3} F_{2}  \\
    & - \frac{47}{15} F_{1}
    F_{2} + \frac{8}{5} F_{2}^{2} + 3 F_{1} F_{0} + 2 F_{0}^{2} \Big)
    + \delta \Big( -2 D_{2}^{2} + 20 D_{3} D_{4} + 5 D_{4}^{2} - \tfrac{7}{4} F_{3}^{2} +
    \tfrac{12}{5} D_{2} F_{1}  \\
    & + \tfrac{11}{10} F_{3} F_{2} - \tfrac{118}{15}
    F_{1} F_{2} + \tfrac{24}{5} F_{2}^{2} + \tfrac{21}{2} F_{1} F_{0} + 7
    F_{0}^{2} \Big).
  \end{split}
\end{equation}
Given $G_4$, one can match the flux-induced Chern-Simons coefficients
with those produced by chiral matter on the F-theory side by solving
\begin{equation}
  \Theta^M_{\Lambda \Sigma} = \Theta^F_{\Lambda \Sigma}. \label{e:CS_match}
\end{equation}
There is a small ambiguity in solving \eqref{e:CS_match} for the
chiral indices, since the two $SU(5)$-singlets give proportional
contributions. We therefore demand that the theory is anomaly-free and
obtain the unique solution
\begin{equation}
  \begin{aligned}
    \chi(\rep{5_{-8}}) &= \alpha\ , \qquad \chi(\rep{5_{-3}}) = \beta\ , \qquad &\chi(\rep{5_{2}}) &= \gamma\ ,  \qquad \chi(\rep{5_{7}}) = \delta\ ,  \\
    \chi(\rep{10_{-1}}) &= -\alpha -\beta - \gamma - \delta \ ,\qquad 
    &\chi(\rep{1_{5}}) &= -8\alpha + \beta + 4 \gamma + 11 \delta \ ,  \\
    \chi(\rep{1_{10}}) &= -5 \alpha + 8 \gamma + 14 \delta\ .  
  \end{aligned}
\end{equation}
Let us stress that it is a non-trivial consistency check that all four
anomaly conditions in Eq.~\eqref{e:su5_anomalies} are satisfied, since
the above ambiguity only gave us one free parameter.

Last of all, we remark that the Euler number $\chi(X_4)$ and
\begin{align}
 \frac{1}{2} \int_{X_4} G_4 \wedge G_4
\end{align}
can easily be calculated using the same toric methods for any choice of flux parameters.
We checked that the tadpole condition can be easily satisfied by constraining the flux 
parameters. Our particular fourfold has $\chi(X_4) = 2364$.

\subsection*{Acknowledgements}

We would like to thank Eran Palti, Raffaele Savelli, Sakura Sch\"afer-Nameki,
and Timo Weigand for interesting discussions and
Dami\'{a}n Kaloni Mayorga Pena for pointing out a mistake in the original
manuscript.
The work of T.G.~and J.K.~was supported by a research grant
of the Max Planck Society. V.B.~was
supported by the Dublin Institute for Advanced Studies. V.B.~would
also like to thank Pas de la Casa for a pleasant surrounding while
this work was being finished.

\appendix

\section{Details on the Calabi-Yau Geometries}

\subsection{Toric Construction}

In the main text we listed the rays of the fan defining the threefold
in \autoref{tab:X3}, and for the fourfold in Eq.~\eqref{eq:X4rays}. To
uniquely determine the ambient toric variety (and, therefore, the
Calabi-Yau hypersurface) it is neccessary to also list the generating
cones of the fan.  Different choices for the fan will result in
different intersection numbers, but not in different $U(1)$ charges.
For the threefold hypersurface, we pick
 \begin{equation} \label{fanSigma}
    \begin{split}
      \Sigma = \smash{\Big\{}&
      \big\langle \Hhat{0}\Fhat{0}\Hhat{1}\Fhat{1} \big \rangle,
      \big\langle \Hhat{0}\Fhat{2}\Hhat{1}\Fhat{1} \big \rangle,
      \big\langle \Dhat{0}\Hhat{0}\Fhat{3}\Fhat{0} \big \rangle,
      \big\langle \Dhat{0}\Fhat{3}\Fhat{0}\Hhat{1} \big \rangle,
      \big\langle \Hhat{0}\Fhat{3}\Fhat{0}\Hhat{1} \big \rangle, \\&
      \big\langle \Hhat{0}\Fhat{3}\Fhat{2}\Dhat{3} \big \rangle,
      \big\langle \Fhat{3}\Fhat{2}\Dhat{3}\Hhat{1} \big \rangle,
      \big\langle \Hhat{0}\Fhat{3}\Fhat{2}\Hhat{1} \big \rangle,
      \big\langle \Hhat{0}\Dhat{1}\Dhat{2}\Fhat{1} \big \rangle,
      \big\langle \Hhat{0}\Fhat{2}\Dhat{2}\Fhat{1} \big \rangle, \\&
      \big\langle \Hhat{0}\Fhat{2}\Dhat{3}\Dhat{2} \big \rangle,
      \big\langle \Dhat{1}\Hhat{1}\Dhat{2}\Fhat{1} \big \rangle,
      \big\langle \Fhat{2}\Hhat{1}\Dhat{2}\Fhat{1} \big \rangle,
      \big\langle \Fhat{2}\Dhat{3}\Hhat{1}\Dhat{2} \big \rangle,
      \big\langle \Dhat{0}\Hhat{0}\Dhat{1}\Fhat{1} \big \rangle, \\&
      \big\langle \Dhat{0}\Hhat{0}\Fhat{0}\Fhat{1} \big \rangle,
      \big\langle \Dhat{0}\Hhat{0}\Dhat{1}\Dhat{2} \big \rangle,
      \big\langle \Dhat{0}\Fhat{0}\Hhat{1}\Fhat{1} \big \rangle,
      \big\langle \Dhat{0}\Dhat{1}\Hhat{1}\Fhat{1} \big \rangle,
      \big\langle \Dhat{0}\Dhat{1}\Hhat{1}\Dhat{2} \big \rangle, \\&
      \big\langle \Hhat{0}\Fhat{3}\Dhat{3}\Dhat{4} \big \rangle,
      \big\langle \Dhat{0}\Hhat{0}\Fhat{3}\Dhat{4} \big \rangle,
      \big\langle \Dhat{0}\Fhat{3}\Hhat{1}\Dhat{4} \big \rangle,
      \big\langle \Fhat{3}\Dhat{3}\Hhat{1}\Dhat{4} \big \rangle,
      \big\langle \Dhat{0}\Hhat{0}\Dhat{2}\Dhat{4} \big \rangle, \\&
      \big\langle \Hhat{0}\Dhat{3}\Dhat{2}\Dhat{4} \big \rangle,
      \big\langle \Dhat{3}\Hhat{1}\Dhat{2}\Dhat{4} \big \rangle,
      \big\langle \Dhat{0}\Hhat{1}\Dhat{2}\Dhat{4} \big \rangle 
      \smash{\Big\}}
    \end{split}
\end{equation}
and for the fourfold hypersurface we are using
 \begin{equation} \label{fanSigma2}
    \begin{split}
      \Sigma = \smash{\Big\{}&
\langle \Hhat{0}\Hhat{1}\Hhat{2}\Fhat{3}\Fhat{2} \rangle, \langle
\Hhat{0}\Hhat{1}\Hhat{2}\Fhat{3}\Fhat{0} \rangle, \langle
\Hhat{0}\Hhat{1}\Hhat{2}\Fhat{1}\Fhat{2} \rangle, \langle
\Hhat{0}\Hhat{1}\Hhat{2}\Fhat{1}\Fhat{0} \rangle,\\& \langle
\Hhat{0}\Hhat{1}\Dhat{0}\Dhat{1}\Dhat{2} \rangle, \langle
\Hhat{0}\Hhat{1}\Dhat{0}\Dhat{1}\Fhat{1} \rangle, \langle
\Hhat{0}\Hhat{1}\Dhat{0}\Dhat{2}\Dhat{4} \rangle, \langle
\Hhat{0}\Hhat{1}\Dhat{0}\Dhat{4}\Fhat{3} \rangle,\\& \langle
\Hhat{0}\Hhat{1}\Dhat{0}\Fhat{3}\Fhat{0} \rangle, \langle
\Hhat{0}\Hhat{1}\Dhat{0}\Fhat{1}\Fhat{0} \rangle, \langle
\Hhat{0}\Hhat{1}\Dhat{1}\Dhat{2}\Fhat{1} \rangle, \langle
\Hhat{0}\Hhat{1}\Dhat{2}\Dhat{3}\Dhat{4} \rangle,\\& \langle
\Hhat{0}\Hhat{1}\Dhat{2}\Dhat{3}\Fhat{2} \rangle, \langle
\Hhat{0}\Hhat{1}\Dhat{2}\Fhat{1}\Fhat{2} \rangle, \langle
\Hhat{0}\Hhat{1}\Dhat{3}\Dhat{4}\Fhat{3} \rangle, \langle
\Hhat{0}\Hhat{1}\Dhat{3}\Fhat{3}\Fhat{2} \rangle,\\& \langle
\Hhat{0}\Hhat{2}\Dhat{0}\Dhat{1}\Dhat{2} \rangle, \langle
\Hhat{0}\Hhat{2}\Dhat{0}\Dhat{1}\Fhat{1} \rangle, \langle
\Hhat{0}\Hhat{2}\Dhat{0}\Dhat{2}\Dhat{4} \rangle, \langle
\Hhat{0}\Hhat{2}\Dhat{0}\Dhat{4}\Fhat{3} \rangle,\\& \langle
\Hhat{0}\Hhat{2}\Dhat{0}\Fhat{3}\Fhat{0} \rangle, \langle
\Hhat{0}\Hhat{2}\Dhat{0}\Fhat{1}\Fhat{0} \rangle, \langle
\Hhat{0}\Hhat{2}\Dhat{1}\Dhat{2}\Fhat{1} \rangle, \langle
\Hhat{0}\Hhat{2}\Dhat{2}\Dhat{3}\Dhat{4} \rangle,\\& \langle
\Hhat{0}\Hhat{2}\Dhat{2}\Dhat{3}\Fhat{2} \rangle, \langle
\Hhat{0}\Hhat{2}\Dhat{2}\Fhat{1}\Fhat{2} \rangle, \langle
\Hhat{0}\Hhat{2}\Dhat{3}\Dhat{4}\Fhat{3} \rangle, \langle
\Hhat{0}\Hhat{2}\Dhat{3}\Fhat{3}\Fhat{2} \rangle,\\& \langle
\Hhat{1}\Hhat{2}\Dhat{0}\Dhat{1}\Dhat{2} \rangle, \langle
\Hhat{1}\Hhat{2}\Dhat{0}\Dhat{1}\Fhat{1} \rangle, \langle
\Hhat{1}\Hhat{2}\Dhat{0}\Dhat{2}\Dhat{4} \rangle, \langle
\Hhat{1}\Hhat{2}\Dhat{0}\Dhat{4}\Fhat{3} \rangle,\\& \langle
\Hhat{1}\Hhat{2}\Dhat{0}\Fhat{3}\Fhat{0} \rangle, \langle
\Hhat{1}\Hhat{2}\Dhat{0}\Fhat{1}\Fhat{0} \rangle, \langle
\Hhat{1}\Hhat{2}\Dhat{1}\Dhat{2}\Fhat{1} \rangle, \langle
\Hhat{1}\Hhat{2}\Dhat{2}\Dhat{3}\Dhat{4} \rangle,\\& \langle
\Hhat{1}\Hhat{2}\Dhat{2}\Dhat{3}\Fhat{2} \rangle, \langle
\Hhat{1}\Hhat{2}\Dhat{2}\Fhat{1}\Fhat{2} \rangle, \langle
\Hhat{1}\Hhat{2}\Dhat{3}\Dhat{4}\Fhat{3} \rangle, \langle
\Hhat{1}\Hhat{2}\Dhat{3}\Fhat{3}\Fhat{2} \rangle
      \smash{\Big\}}\;.
    \end{split}
\end{equation}
We note that these fans have been constructed such that they are
compatible with the projection that we want to use as a toric
morphism.

\subsection{Showing the Existence of a Section} \label{app:showsection}

In this appendix we show that the homology class \eqref{eq:MWsection} actually 
contains a section, and does not only happen to have the right intersection 
numbers with fiber components.
To settle this question we have to compute the line bundle cohomology
group $H_0(X,\Osheaf_X(s))$. This cohomology group sits in the long
exact sequence for the sheaf exact sequence
\begin{equation}
  0 \longrightarrow
  \Osheaf_{X_\Sigma}(s+K_{X_\Sigma})  \longrightarrow
  \Osheaf_{X_\Sigma}(s) \longrightarrow
  \Osheaf_X(s) \longrightarrow
  0
\end{equation}
for the restriction from the four-dimensional ambient toric variety to
the three-dimensional Calabi-Yau hypersurface. The toric cohomology
groups can easily be computed to be 
\begin{equation}
  \dim H^i\big(X_\Sigma, \Osheaf_{X_\Sigma}(s+K_{X_\Sigma})\big) =
  \begin{cases}
    1 & i=1,\\
    0 & \text{else},
  \end{cases}
  \qquad
  \dim H^i\big(X_\Sigma, \Osheaf_{X_\Sigma}(s)\big) = 0.
\end{equation}
Therefore, the long exact sequence
\begin{multline}
  \cdots \longrightarrow
  H^0(X_\Sigma, \Osheaf_{X_\Sigma}(s)) \longrightarrow
  H^0(X,\Osheaf_X(s)) \longrightarrow
  \\ \longrightarrow
  H^1(X_\Sigma, \Osheaf_{X_\Sigma}(s+K_{X_\Sigma})) \longrightarrow
  H^1(X_\Sigma, \Osheaf_{X_\Sigma}(s)) 
  \longrightarrow \cdots
\end{multline}
tells us that the homology class $[\sigma_1]=
[\Fhat{1}-\Fhat{0}-\Dhat{0}-\Dhat{3}-\Dhat{4}+\Hhat{0}]$ contains a
unique variety $\sigma_1$ representing it.


\bibliography{Jan,Thomas,Volker}
\bibliographystyle{utcaps}

\end{document}